\documentclass[letterpaper]{ifacmtg}
\usepackage[dcucite]{harvard}
\usepackage{epsfig}
\usepackage{amssymb}
\usepackage{graphicx}
\begin{document}
\runauthor{Saifullah}
\begin{frontmatter}
\title{FEEDBACK CONTROL OF PROBABILITY AMPLITUDES FOR TWO-LEVEL ATOM IN OPTICAL FIELD}
\author[Pakistan]{Saifullah}
\address[Pakistan]{School of Mathematical Sciences\\Government College University\\Lahore -- Pakistan\\Email: saifullahkhalid75@yahoo.com}
\begin{abstract}
We demonstrate the possibility to stabilize the probability amplitude of the upper level
for a single quantum two-level atom in a classical optical field with
 feedback control scheme.
\end{abstract}
\end{frontmatter}

\section{Introduction}
The methods of feedback control are widely used in the modern
physics, but still they are not very popular in quantum optics. Very
often this "cybernetical" approach does not demand involvement in
very complicated physical devices and can be arranged in a trivial
nonlinear system [1].

We apply this technique to control the energy of a two-level atom in
the optical external field $E(t)$ in the frame of the so-called
"semi classical model" of the atom--field interaction that describes
a single quantum two-level atomic system (all other levels are
neglected) with classical electromagnetic field.

Recently other authors studied the control of two-level atoms in the
frame of open loop-ideology when the controlling field was known
{\it a priori}. It allowed obtaining the different forms of atomic
energy spectra, producing $\pi$- and $\frac{\pi}{2}$- pulses [2],
taking special non-constant shapes of external field [3] etc.

The main feature of the model proposed in this article is that, it
is based on the closed-loop approach. This means that we do not
initially define the dependency of the field on time, but restore
this function for every moment from the current values of the
probability amplitudes of the atomic ground and excited levels.

The closed-loop (feedback) scheme for the interaction of two-level
atom with external field can be realized in different models. The
most famous and fully developed is the approach based on master
equations in its both main variants: the Markovian feedback model
[4] and the so called Bayesian feedback [5] (the later model was
proposed by Wiseman in his comparative analysis of both these models
in [6]). The Bayesian ideology is more closely related to our
approach because this closed-loop control is constructed directly on
the estimation of the system state.

Another approach is to construct the control scheme for a single
atom for the quantum control field [7]. In this paper we discuss the
classical control field and we do not apply special restrictions on
its shape i.e, the optical field shouldn't be sinusoidal as in [8]
or have other special time dependency. Thus, our scheme of classical
feedback proposed here is similar to the traditional variant of
control theory in the form of speed-gradient (SG) method [9], when
input variables change proportionally to the speed-gradient of
appropriate goal function. We use the standard notations following
[10], but in our model the optical field plays the role of a control
signal $u(t)$ for closed-loop or feedback control scheme.

For this purpose we use the real positive goal function $Q$,
measuring how far at the moment we are from the desired state of the
atom. As a result, we calculate the control signal $u(t)$, i.e. we
restore the shape of the electromagnetic filed $E(t)$ to keep the
atom at the upper level.

In the second section of this work, we construct the feedback
control model for the single two-level atom in external controlling
optical field. Then, in the third section, we apply feedback speed
gradient scheme to the non-decay case.

\section{Two-level atom in control optical field}
Let's consider the interaction of an optical field $E(t)$ linearly
polarized along the $x$-axis with a two-level atom.\\

\hrule
\begin{center}
\epsfig{file=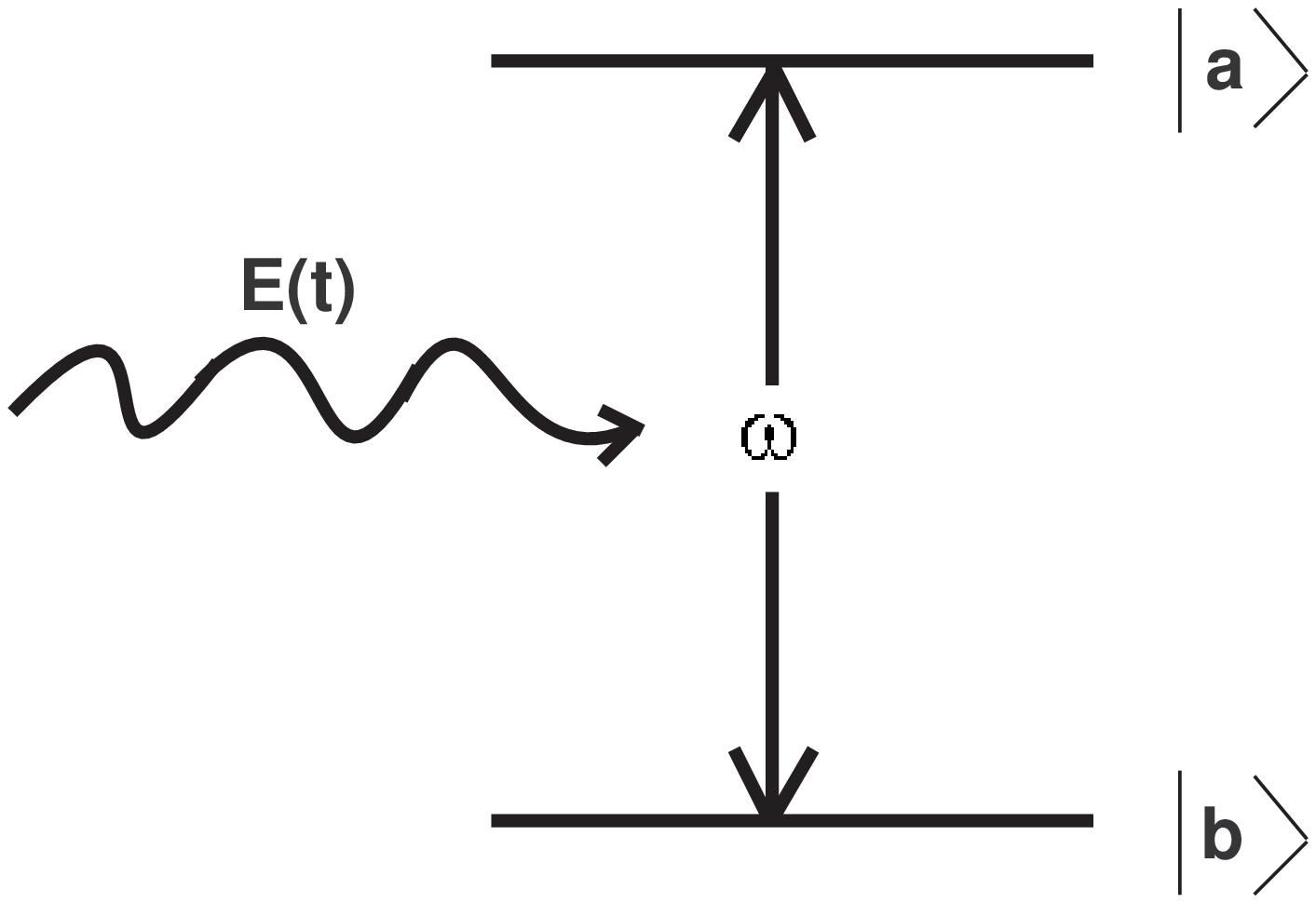,width=6.0cm}
\end{center}
\hrule
\begin{figure}[h]
  \begin{center}
  \caption{Interaction of a single two-level atom with an optical field.}
  \label{}
  \end{center}
\end{figure}

Let $| a\rangle$ and $| b\rangle$ represent the upper and lower
level states of the atom, i.e. they are eigenstates of the
unperturbed part of the Hamiltonian $\hat H_0$ with the eigenvalues:
$\hat H_{0}|a\rangle=\hbar\omega_a|a\rangle$ and $\hat
H_{0}|b\rangle=\hbar\omega_b|b\rangle$. The wave function of a
two-level atom can be written in the form
$$
|\psi(t)\rangle=C_a(t)|a\rangle +C_b(t)|b\rangle,
$$
where $C_a$ and $C_b$ are the probability amplitudes of finding the
atom in states $|a\rangle$ and $|b\rangle$, respectively. The
corresponding Schr\"{o}dinger equation is:
$$
|\dot{\psi(t)}\rangle=-\frac{\iota}{\hbar}\hat
H|\psi(t)\rangle,
$$
with $\hat H=\hat H_{0}+\hat H_1$, where $\hat H_{0}$ and $\hat H_1$
represent the unperturbed and interaction parts of the Hamiltonian,
respectively [10]:
\begin{eqnarray}
\nonumber \hat H_{0}&=&\hbar\omega_a|a\rangle\langle
a|+\hbar\omega_b|b\rangle\langle b| \ ;\\
 \nonumber
\hat H_1&=&-\Big(\wp_{ab}|a\rangle\langle
b|+\wp_{ba}|b\rangle\langle a|\Big)E(t),
\end{eqnarray}
where $\wp_{ab}=\wp^*_{ba}=e\langle a|x|b\rangle$ is the matrix
element of the electric dipole moment. We neglect the decay of the
levels. We express the electric field as
$$
E(t)=E_0 u(t),
$$
where $E_0$ is the amplitude and $u(t)$ is the dimensionless control
signal. The equations of motion for the
amplitudes $C_a$ and $C_b$ may be written as
$$
\dot{C_a}=-\iota \omega_a C_a+\iota\Omega_R u(t)e^{-\iota\phi}C_b;
$$
$$
\dot{C_b}=-\iota \omega_b C_b+\iota\Omega_R u(t)
e^{\iota\phi}C_a,
$$
where the "Rabi frequency" is defined as $\Omega_R
=\frac{|\wp_{ba}|E_{0}}{\hbar}$, and $\phi$ is the phase of the
dipole matrix element $\wp_{ba}=|\wp_{ba}|e^{\iota\phi}$.

To solve for $C_a$ and $C_b$, we write the equations of motion for
the slowly varying amplitudes as:
$$
c_a = C_a e^{\iota \omega_a t}\ \ ; \ \ c_b = C_b e^{\iota
\omega_b t},
$$
then
\begin{eqnarray}
\nonumber
\dot{c_a}=\iota\Omega_R u(t)e^{-\iota\phi}c_b e^{\iota
\omega t}\ ;\\
\nonumber \dot{c_b}=\iota\Omega_R u(t)e^{\iota\phi}c_a e^{-\iota
\omega t}\ ,
\end{eqnarray}
where $\omega = \omega_a - \omega_b$ is the atomic transition
frequency. The phase $\phi$ can be excluded from the system, if we
put $\widetilde{c_b}=c_be^{-\iota\phi}$:
$$
\dot{\widetilde{c_b}}=\iota\Omega_R
u(t)e^{-\iota w t}c_a
$$

Later for simplicity we will denote $\widetilde{c_b}$ with $c_b$, then finally:
\begin{equation}
\label{16}
\dot{c_a}=\iota\Omega_R u(t)e^{\iota
w t}c_b
\end{equation}
\begin{equation}
\label{17}
\dot{c_b}=\iota\Omega_R u(t)e^{-\iota
w t}c_a
\end{equation}
Now let's suppose that we have the initial conditions:
\begin{equation}
\label{ic}
c_a(0)=0\ \ ;\ \ c_b(0)=1
\end{equation}
and our goal is to stabilize the atomic system at the upper level:
$|c_a|^2=1$.

\section{Speed gradient method for probability amplitudes control}
We have not yet specified the time-dependent function $u(t)$. To
find it, we apply the speed gradient (SG) method [9] to control the
system behavior.

In this approach, the control action is chosen in the maximum
descent direction for a scalar goal function.

The goal in the control process is a smooth scalar function $Q$ with the
limit relation $$\lim_{t\rightarrow\infty}Q(x(t),t)\rightarrow 0.$$
The purpose of the SG method is to minimize the goal function
\begin{equation}
\label{goal}
Q= \frac{1}{2}\Big(|c_a|^2-1\Big)^2,
\end{equation}
where $|c_a|^2=c_a c^*_a$.

SG represents the control signal $u$ with the time derivative of the
goal function $Q$.

The underlying idea of SG method is that moving along the
anti-gradient of the speed $\dot Q$ provides decreasing of the goal
function. In our case the control signal space is 1-dimensional,
thus we reduce our gradient to the partial derivative with respect
to $u$. In the case of proportional feedback with some positive
coefficient $\Gamma$, it is defined in the form:
\begin{equation}
\label{sg} u=-\Gamma\frac{\partial\dot{Q}}{\partial u}
\end{equation}
Thus
\begin{eqnarray}
\label{18} u(t)=\iota\Gamma\Omega_R\Big(|c_a|^2
-1\Big)\Big(e^{-\iota wt}c_a c^*_b - e^{\iota wt}c_b
c^*_a\Big)
\end{eqnarray}
Putting value of $u(t)$ from Eq.(\ref{18}) in Eqs.(\ref{16}) and
(\ref{17}), we have the following system of equations:
\begin{eqnarray}
\nonumber \dot c_a=\Gamma\Omega^2_R\Big(|c_a|^2
-1\Big)\Big(e^{2\iota wt}c^*_a
c^2_b-c_a|c_b|^2 \Big) \ ;\\
\nonumber \dot c_b =\Gamma\Omega^2_R\Big(|c_a|^2 -1\Big)\Big(c_b
|c_a|^2 -e^{-2\iota wt} c^2_a c^*_b\Big) .
\end{eqnarray}
Now suppose that
\begin{eqnarray}
\nonumber \rho_a=c_a c^*_a=|c_a|^2 \ \ ; \ \
 \rho_b=c_b c^*_b=|c_b|^2 \ \ ;\\
\iota \rho_- = e^{-\iota\omega t}c_a c^*_b - e^{\iota\omega
t}c^*_a\nonumber c_b \ \ ;\\ \rho_+ = e^{-\iota\omega t}c_a c^*_b +
e^{\iota\omega t}c^*_a c_b.\nonumber
\end{eqnarray}
Hence we have the following four equations:
\begin{eqnarray}
\nonumber \dot{\rho_a}=2\Gamma\Omega^2_R\Big(\rho_a
-1\Big)\Big[\Big(\frac{\rho^2_+ - \rho^2_-}{4}\Big)-\rho_a\rho_b\Big] \ ;\\
\nonumber
  \dot{\rho_b}=2\Gamma\Omega^2_R\Big(\rho_a
  -1\Big)\Big[\rho_a\rho_b-
\Big(\frac{\rho^2_+ - \rho^2_-}{4}\Big)\Big] \ ; \\
\label{cc3}
\dot{\rho_+}=\omega \rho_- \ ; \\
\nonumber
  \dot{\rho_-}=-2\Gamma\Omega^2_R\Big(\rho_b -\rho_a\Big)
\Big(\rho_a -1\Big)\rho_--\omega\rho_+ \ .
\end{eqnarray}
Also from Eq.(\ref{18}) the control signal $u(t)$ becomes
\begin{equation}
\label{ufinal} u(t)=-\Gamma\Omega_R \Big(\rho_a -1\Big)\rho_-
\end{equation}

With initial conditions $\rho_a(0)=0, \,\, \rho_b(0)=1$ we have
$$
\dot{\rho_a}+\dot{\rho_b}=0 \ ,
$$
that means in fact:
$$
|c_a(t)|^2 + |c_b(t)|^2 = 1,
$$
which is the simple statement that the probability to find the atom
in one of its states $|a\rangle$ or $|b\rangle$ is 1.

Thus, we can
simplify the system (\ref{cc3}), putting $\rho_b=1-\rho_a$.

The system (\ref{cc3}) has two equilibrium (fixed) points:
$$
(\rho_a, \rho_+, \rho_-)=(0, 0, 0 ), (1, 0, 0)
$$

On the Figs. 2,3 we demonstrate the result of our control procedure
for: $\Gamma =0.1 \ sec$, $\Omega _R=10^2 \ sec^{-1}$ and
$\omega=10^3 \ sec^{-1}$.

On Fig.2 we show the solution of Eq.(\ref{cc3}a). \\
\hrule
\begin{center}
\epsfig{file=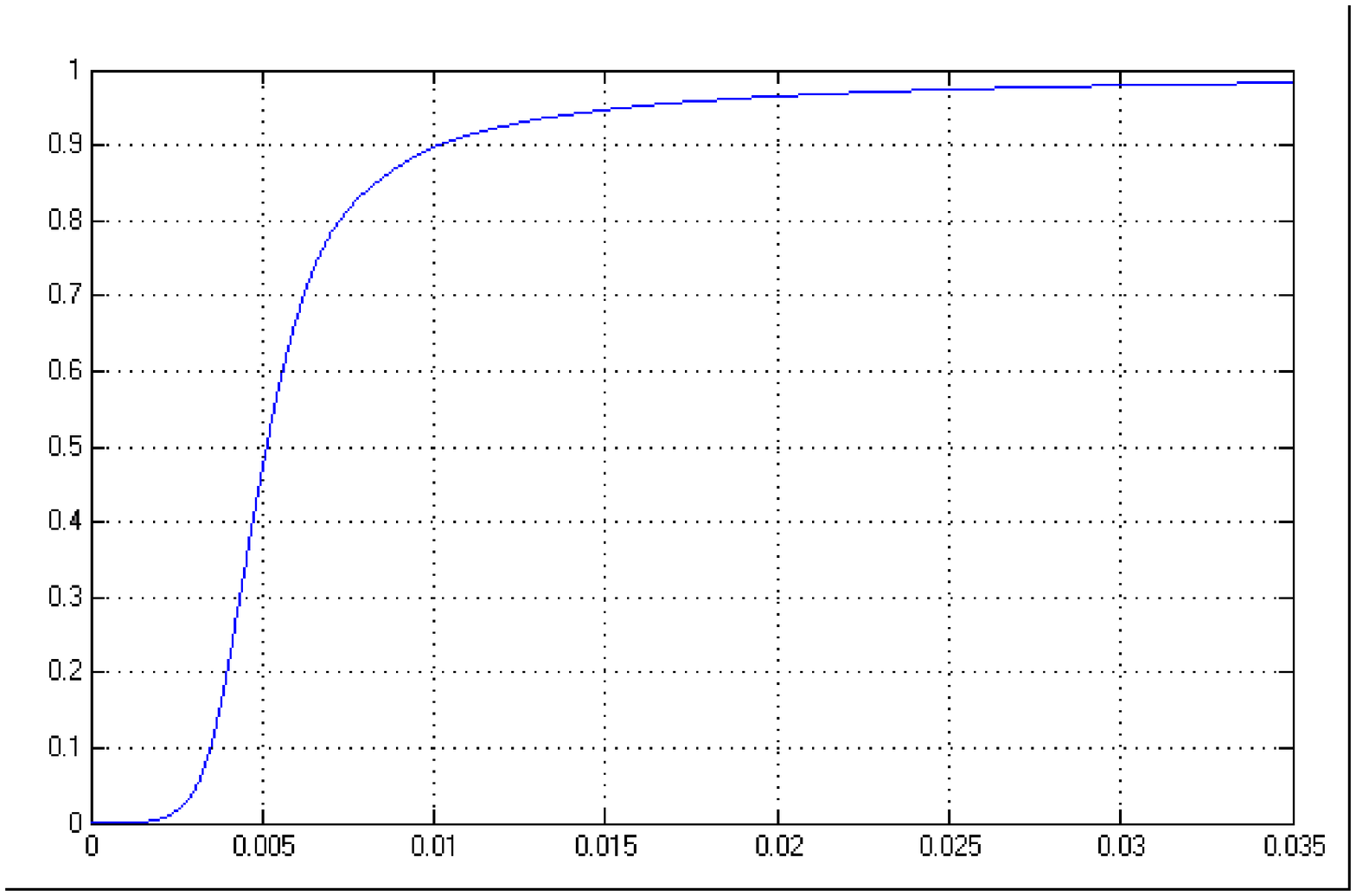,width=7.0cm}
\end{center}
\hrule
\begin{figure}[h]
  \begin{center}
   \caption{The density matrix element $\rho_a (t)$ for the control procedure (\ref{goal})-(\ref{sg})}.
  \label{}
  \end{center}
\end{figure}

On Fig.3 we show the control signal(\ref{ufinal}).\\
\hrule
\begin{center}
\epsfig{file=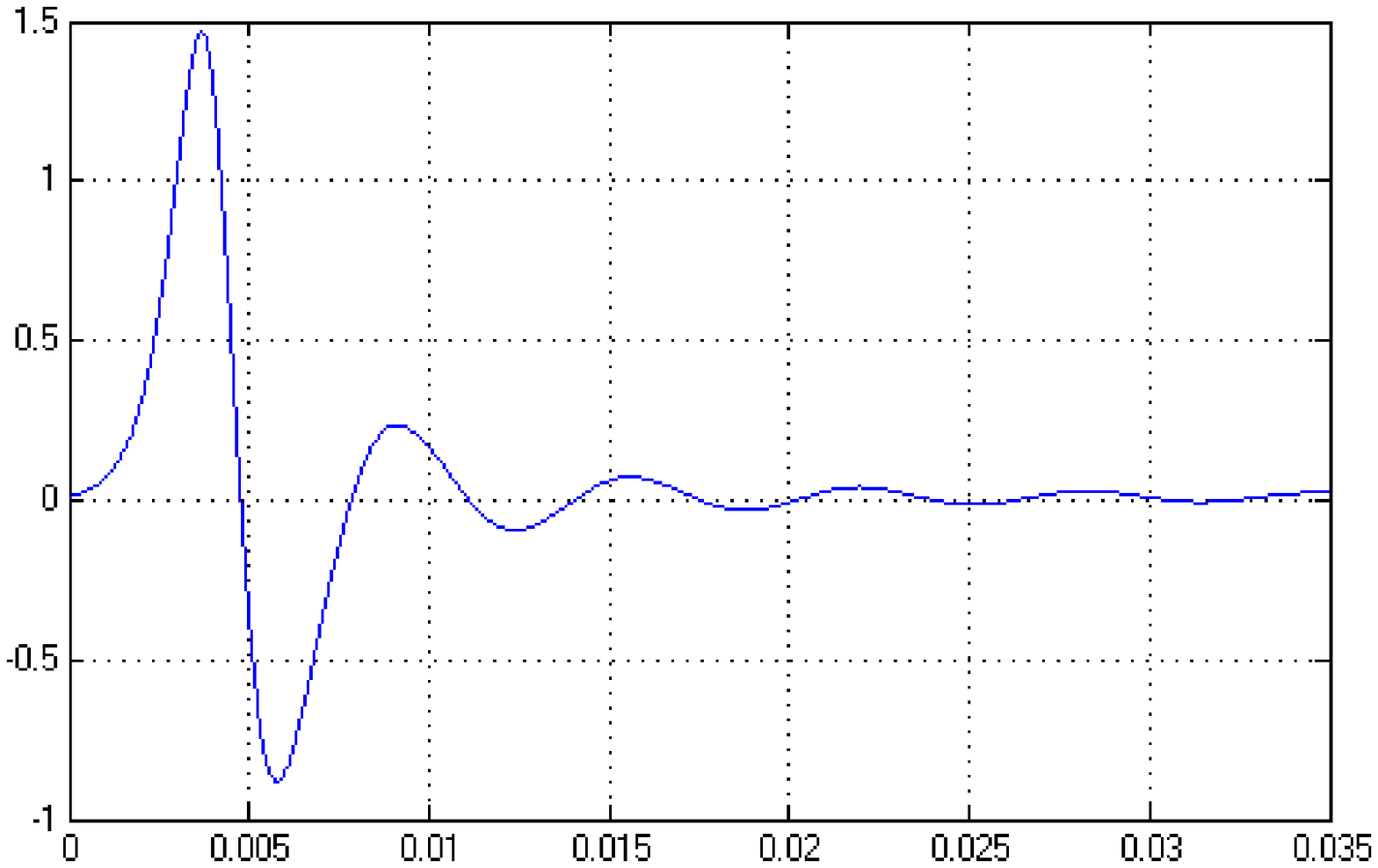,width=7.0cm}
\end{center}
\hrule
\begin{figure}[h]
  \begin{center}
  \caption{The control signal $u(t)$ for the system (\ref{cc3}).}
  \label{}
  \end{center}
\end{figure}

\section{Conclusion}
The SG algorithm can be easy applied to establish feedback control
for the probability amplitudes of two-level atom.

This scheme can be modified if we take into consideration the decay
of the atom levels, because in this case the goal $Q =
\frac{1}{2}\Big(|c_a|^2 - 1 \Big)^2$ is not achievable for SG
algorithm in principle. For this purpose we will redefine the goal
function $Q$.

\section{Acknowledgement}
I wish to thank the reviewer(s) for their valuable comments and
suggestions, which significantly improved the paper.

I am also thankful to Dr. Sergei Borisenok (Department of Physics,
Herzen State University, Saint Petersburg, Russia) for productive
scientific discussions, valuable suggestions and guidance.

\end{document}